\documentclass[aps,prb,reprint]{revtex4-1}
\usepackage{lipsum}
\usepackage{amssymb,amsmath}
\usepackage{subfig}
\usepackage{graphicx}
\usepackage{color}
\usepackage{hyperref}
\usepackage{listings}
\usepackage{fancybox}
\usepackage{caption}
\makeatletter
\newcommand*{\balancecolsandclearpage}{%
  \close@column@grid
  \cleardoublepage
  \twocolumngrid
}

\makeatother
\begin{document}

\title{Single-cell unroofing: probing topology and nanomechanics of native membranes}
\author{Nicola Galvanetto}

\affiliation{International School for Advanced Studies (SISSA), via Bonomea 265, Trieste 34136, Italy; Email: \texttt{nicola.galvanetto@sissa.it}}

\date{Published in \textit{Biochimica et Biophysica Acta} (2018); DOI: \url{https://doi.org/10.1016/j.bbamem.2018.09.019}}

\begin{abstract}
	Cell membranes separate the cell interior from the external environment. They are constituted by a variety of lipids; their composition determines the dynamics of membrane proteins and affects the ability of the cells to adapt. Even though the study of model membranes allows to understand the interactions among lipids and the overall mechanics, little is known about these properties in native membranes. To combine topology and nanomechanics analysis of native membranes, I designed a method to investigate the plasma membranes isolated from a variety of single cells. Five cell types were chosen and tested, revealing 20\% variation in membrane thickness. I probed the resistance of the isolated membranes to indent, finding their line tension and spreading pressure. These results show that membranes isolated from neurons are stiffer and less diffusive than brain cancer cell membranes. This method gives direct quantitative insights on the mechanics of native cell membranes.
	
\end{abstract}

\maketitle

\section{Introduction}
When the first observation of the bi-layered nature of cell membrane was made using the electron microscope in 1959 [1], the notion that it was composed of lipid layers had already been accepted for decades  [2,3]. Since then, much has been discovered in terms of membrane composition and function, mostly thanks to biochemical approaches.
The direct imaging of a native membrane remained challenging for years because of its fluid nature. Major advancements occurred only in the Seventies when various techniques showed their potential and gave birth to what is now broadly called ‘cell unroofing’. 
They involved the separation of cell cortices by violent mechanical treatments [4], combined with electron microscopy. To observe the membrane isolated from the cytosolic environment, the cells were firstly deposited on a surface coated with a ``glue`` (e.g. poly-L-lysine or Alcian blue), and then broken. The aim is to isolate the cell membrane adherent with the substrate from the rest of the cell. Three are the main strategies that can be applied. The first strategy is to expose the cells to a strong lateral flux of medium: this will break the cells leaving residues of membranes attached to the substrate [5].  The second is fracturing: it consists on sandwiching the cells, freeze and separate the sandwich [6,7]; this allowed to achieve a more natural, life-like appearance of the samples. A variant of the fracturing method is the recently developed iMEM [8] that consist of isolating the cell membranes during the blotting step of a Cryo-EM grid preparation. The third uses sonic waves to break the body of the cells: in this way only the layer of membrane in contact with the substrate remains [9]. The preparation usually ends with a physical fixation of the sample and the investigation with the electron microscope. 
Looking directly to the cytosolic side of the membrane expanded our knowledge on the internal architecture of the cell membrane that before was simply not accessible [10,11].
Another natural instrument of investigation with sub-nanometer precision is the atomic force microscope (AFM). AFM has a big advantage compared to light or electron microscopy, in fact the sample doesn’t require neither fluorescent labels nor metal coating or cryo-fixation. Some recent studies [12-16] unraveled a detailed architecture of the cytoskeleton on the inner face of fixed membranes with a resolution of \~5nm. 
However, the preparation of these samples requires a considerable know-how to be successful and to become reproducible in large scale [13]; moreover the aforementioned treatments act on the entire cultures without any fine control of the process. In fact, the studies that required the manipulation of lipid membranes and vesicles concentrated mostly on synthetic  preparations of mixtures of lipids [17-19] . Model membranes are in particular suitable for the study of lipids interactions [20-23] and to test how the membranes behave in their physiological environment. These studies demonstrated that the strength of the bilayer is enhanced by lowering the temperature [24,25], or when exposed to high ion concentrations [26] or in a mixture with cholesterol that mimic lipid rafts [27,28]. 
Here I present a method that allows the direct study of native membranes in buffer solution and under ambient pressure and temperature. I present an AFM topography analysis that depicts heights differences in the membrane of different cell types. I measured their breaking forces which allowed to determine the line tension and the spreading pressure, through the indentation of the bilayers with the tip of the AFM. Furthermore, I discuss future improvements to obtain sub-molecular resolution of membrane proteins in their native environment and the requirements for in vivo applications.

\begin{figure*}
	\centering
	\includegraphics[width=0.8\linewidth]{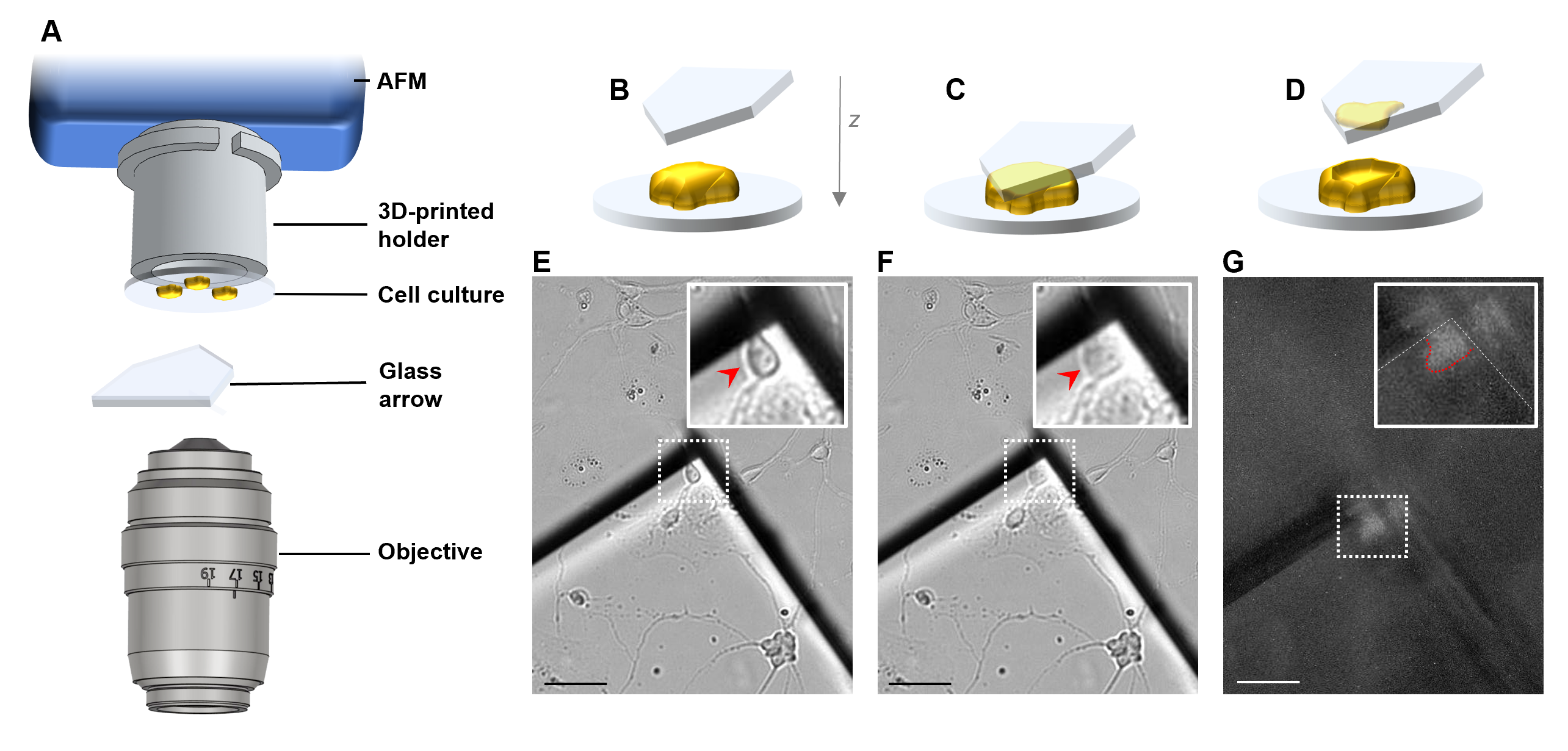}
	\captionsetup{justification=raggedright,singlelinecheck=false}
	\caption{(A) Scheme of the apparatus: the cell culture holder is mounted on the head-stage of the AFM which moves the cell culture downward against the fixed glass arrow. The process is controlled using the optical microscope.  (B-C-D) Exemplificatory draws of the compression of one cell (the scheme is upside-down, and the cell is enlarged for clarity). (E-F-G) images of the compression of the target cell. (B-E) before compression, (C-F) during compression the cell is squeezed, (D-G) after compression. (G) The cell was stained with a fluorescent membrane dye (DII) to identify the membrane with the fluorescent image. }
	\label{fig:cpu}
\end{figure*}

\section{Materials and Methods}

\subsection{Overview of the setup for sample preparation}

The setup was based on an AFM (JPK Nanowizard III) mounted on an inverted optical microscope (Olympus IX71) as sketched in Figure 1A and Figure S1. The three central elements of Figure 1A consist of a cell culture coverslip, mounted on a holder, and squeezed with a triangular glass (see the next Sections for the details). During the sample preparation process, the AFM was used just as a motor with micrometer precision, indeed it was obtained the same results squeezing the cells with a water three-axis micromanipulator.

\subsection{Cell cultures}

The method was tested with 5 cell types: U87, U251, HEK293, primary hippocampal neurons, primary dorsal root ganglia (DRG) neurons. 
The preparation of the glass coverslip was identical for all the five cell types: glass round coverslips (12 mm in diameter, 200 $\mu$m in thickness) were plasma cleaned for 15 seconds, coated with 0.5 mg/ml poly-D-lysine (Sigma-Aldrich, St. Louis, MO, USA) for 1 h at 37 Celsius and washed 3 times in deionized water. The coating is necessary to create a strong adhesion between the cells and the substrate because they must resist the compression and the laceration.
Hippocampal and DRG neurons from Wistar rats (P2-P3) were prepared in accordance with the guidelines of the Italian Animal Welfare Act, and their use was approved by the Local Veterinary Service, the SISSA Ethics Committee board and the National Ministry of Health (Permit Number: 2848-III/15) in accordance with the European Union guidelines for animal care (d.l. 26, March 4th 2014 related to 2010/63/UE and d.1. 116/92; 86/609/C.E.). The animals were anesthetized with CO2 and sacrificed by decapitation, and all efforts were made to minimize suffering. Dissociated cells were plated at a concentration of 4 × 104 cells/ml. The medium used for hippocampal neurons is in Minimum Essential Medium (MEM) with GlutaMAX supplemented with 10\% Fetal Bovine Serum (FBS, all from Invitrogen, Life Technologies, Gaithersburg, MD, USA), 0.6\% D-glucose, 15 mM Hepes, 0.1 mg/ml apo-transferrin, 30 $\mu$g/ml insulin, 0.1 $\mu$g/ml D-biotin, 1 $\mu$M vitamin B12 (all from Sigma-Aldrich), and 2.5 $\mu$g/ml gentamycin (Invitrogen). The medium used for DRG neurons is Neurobasal medium (Gibco, Invitrogen, Milan, Italy) supplemented with 10\% Fetal Bovine Serum (FBS, from Invitrogen, Life Technologies, Gaithersburg, MD, USA). The experiments were performed from two to five days after dissociation.
The human HEK293, U87 and U251 cell lines were cultured with Dulbecco’s modified Eagle’s media (DMEM, Gibco) supplemented with antibiotics (100 U/ml penicillin, 100 $\mu$g/ml streptomycin, Gibco), GlutaMax Supplement (2 mM, Gibco), and 10\% fetal bovine serum (Biowest, USA). All the cells were grown under standard culture conditions (37  Celsius and 5\% CO2).

\subsection{Cell culture holder}

The scope of the cell culture holder is to connect the cell culture (i.e. a coverslip) to a device that moves it with micrometer precision. The holder was designed for an AFM JPK Nanowizard III mounted on an Olympus IX71 optical microscope (but previous trials were successfully performed with a water three-axis manipulator). In Figure S1 is shown the comparison between the original cantilever holder and the cell culture holder: the two share the same bayonet couplings. The cell culture holder is a cylinder that allows to glue the coverslip with Vaseline in the base of the cylinder. The cell culture holder was drawn in Solidworks and printed in resin (visijet m3 black) with a Projet 3510 HD 3D system (see Figure S1 for more details; for the Solidworks file please contact the corresponding author). With the adaptation of the bayonet couplings, an equivalent coverslip holder can be used on every AFM-Inverted microscope setup. 

\subsection{Triangular coverslips preparation (Glass arrows) }

Glass coverslips (24 mm in diameter, 200 $\mu$m in thickness) were plasma cleaned for 15 seconds and broken in 4 quarter with the use of the hands. This passage is crucial and better explained in Figure S2. It is worth noting that the coverslips cannot be cut with a diamond tip because the fracture must be sharp. The resulting quarters (called glass arrows) was immersed in 0.5 mg/ml poly-D-lysine (Sigma-Aldrich, St. Louis, MO, USA) for 30 minutes. The glass arrows were immersed in deionized water for 10 seconds before use.

\subsection{Cell squeezing}

The cell squeezing was performed bringing in contact two parts: the arrow and the cell culture (Figure 1B-G). 
Lower part: tilted arrow. The cover of a petri dish was filled with Ringer solution (2 ml). The arrow was placed tilted of 7-15 degrees (see Figure S1) in the middle of the cover. The cover of the petri dish was then fixed on the stage of the AFM. 
Top part: the cell culture was glued with Vaseline to the holder, and they were mounted on the AFM head stage (Figure S1).  The Head Stage was put on top of the AFM in measurement position. Gradually, the cell culture was immersed into the solution, by lowering the head stage with the electrical motors of the AFM. The distance between the cell culture and the arrows was controlled with the help of the optical microscope. Once the distance reached \~50 $\mu$m, I chose the target cell, centering the apex of the arrow. I set the focus of the microscope in the position of the arrow and I brought the cell in contact with the arrow 2 $\mu$m at a time. When the cell touches the arrow, it enlarges. I continued to approach until the cell doubled its area. I kept it in contact for 3 minutes and then I rapidly lifted the coverslip. Nothing, or just few cellular debris can be visible in the contact region, otherwise the preparation will result contaminated. The arrow was then laid down and fixed (see Figure S1 D and E). The medium was replaced slowly, without drying completely the solution (the isolated membrane should not come out of the solution).

\subsection{AFM Imaging and Force Spectroscopy}

AFM imaging and force spectroscopy were performed with a Nanowizard III system (JPK) mounted on an inverted optical microscope (Olympus IX71), using Hydra NGG (Appnano) cantilevers with nominal spring constant of 0.084 N/m. AFM images were taken in intermittent contact mode applying the lowest possible force during imaging (cantilever free oscillation amplitude was set at 18-20 nm, the surface was approached and scanned at \~75\% of the free amplitude). Force-distance curves were acquired at a rate of 2000 nm/s. AFM imaging and force spectroscopy were performed in Ringer’s solution (NaCl 145 mM, KCl 3 mM, CaCl2 1.5 mM, MgCl2 1 mM, Glucose 10 mM, HEPES 10 mM adjusted to pH 7.4 with NaOH) at 24 Celsius. The AFM height sensor was calibrated with the TGZ01 calibration grid before and after the series of experiments.

\subsection{Data analysis}

 The AFM images were processed with the open source software Gwyddion [29]. The height measure (for each topographical image) was calculated as described in Figure S3: briefly, the resulting height of the membrane is here defined as the difference between the mean of the z-position of the substrate and the mean of the z-position of the membrane (\~10,000 z-positions for each level). Then, the mean of three height measures for each cell type is computed and reported in Fig. 2F (n=\~10,000/3 z-positions/samples). The error bars represent the standard error of the mean. The roughness was defined as the root mean square (RMS) of the height distribution, and calculated in Gwyddion. The force-distance curves and the force maps were analyzed with a modified version of the open source software Fodis [30] that automatically detects the baseline and the Breakthrough force. Breakthrough force averages of 800-1000 force spectra were calculated for each cell type and reported in Fig. 3B. The error bars represent the standard deviation of the distributions. The nucleation model was applied to the breaking force distributions and fitted with an add-on module of Fodis following the procedure described by Chiantina et al. [31]. The modified version of Fodis is available at \url{https://github.com/nicolagalvanetto/Fodis}.

\section{Results and Discussion}

\subsection{Upper membrane separation}

I developed a method that allows the combination of topographical and mechanical investigations of native cell membranes. For this, I needed to make a preparation of exposed native membranes, separated from the entire cell, and arranged on a neutral support.
The method was designed to fit two additional requirements: i) the operation had to be simple and reproducible; ii) the membrane had to be localized easily (it is not visible by bright-field microscopy).
For this purpose, I designed a cell culture holder (see Fig 1A and Figure S1) to connect the cells to a motor with micrometer precision, and a new protocol to prepare optically-sharp arrows from common glass coverslips (Figure S2).  The target cell is brought in contact with the apex of the arrow and squeezed for \~3 minutes. Then, the cell culture is rapidly moved away. The membrane that went in contact with the arrow is torn from the cell as a consequence of the interaction between the membrane itself and the polylysine coating of the arrow.
Remarkably, the membrane remains on top of the flat corner of the arrow, like the footprint of the squeezed cell. The extracted membrane is therefore poised for various investigations, in fact it is exposed to the buffer which can include specific ligands that can activate membrane proteins involved in signal transduction, or in metabolic and bioenergetics processes.
I tested 5 cell types of various shapes and dimensions: one human epithelial cell line (HEK293), two human brain cancer cell lines (U87, U251) and two types of primary neurons (Hippocampal and dorsal root ganglia neurons (DRG) from rats). The proposed method was successful for all but the HEK293. The method fails with cell types that do not adhere well to the cell culture coverslip, so that they remain attached to the arrow after compression, without breaking. Improving the adhesion of the cells to the cell culture will expand the usable cell types for this method. The success rate, i.e. the production of samples suitable for AFM investigations, was \~80\%. 

\begin{figure*}
	
	\centering
	\includegraphics[width=0.8\linewidth]{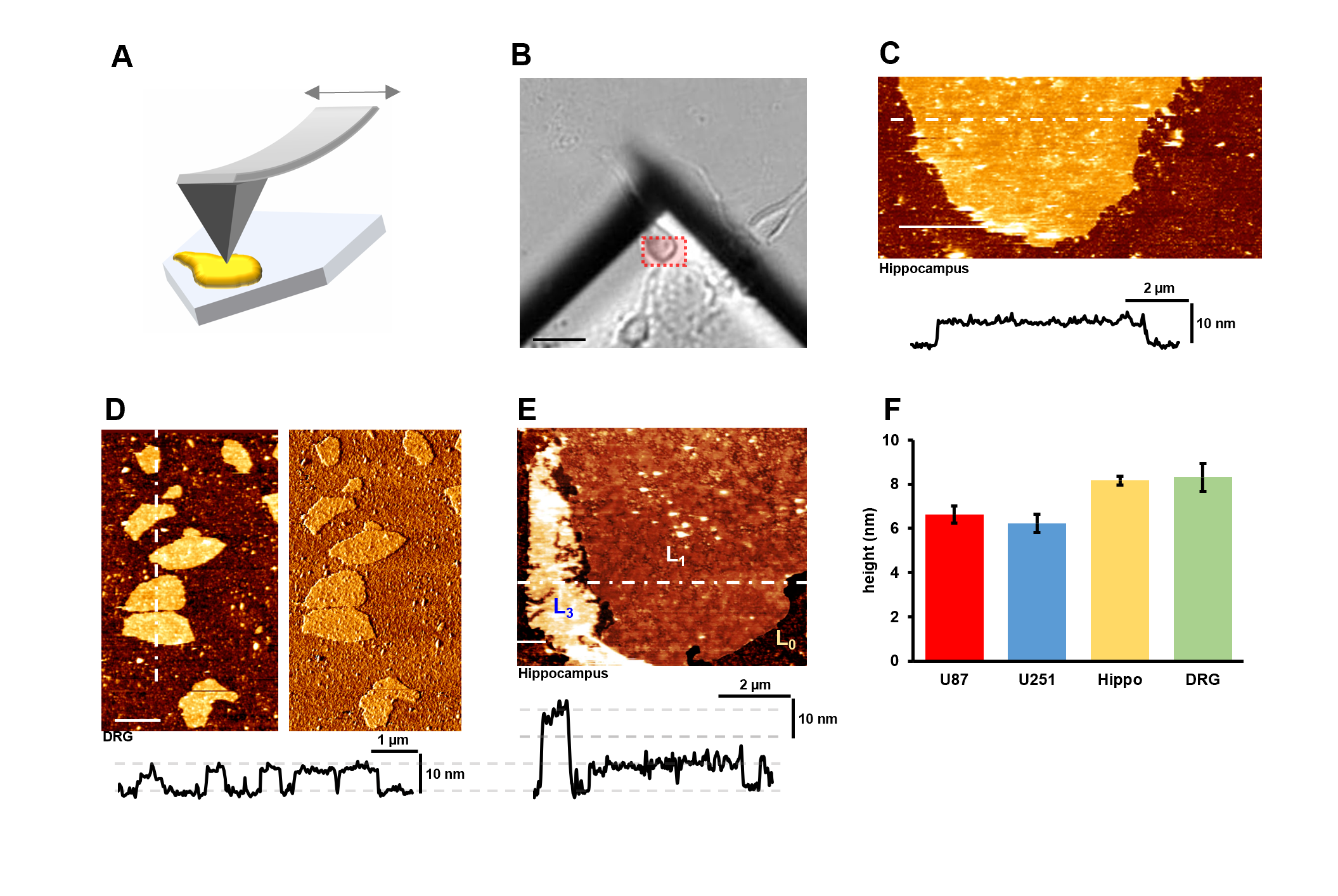}
	\captionsetup{justification=raggedright,singlelinecheck=false}
	\caption{(A) Scheme of the imaging process. (B) Hippocampal neuron during compression, the red area represents the area imaged in (C) on the glass arrow, after the removal of the cell culture (scalebar 15 µm). (C) AFM topography image of the upper membrane of a hippocampal neuron (scalebar 4 µm) and the z-profile under the dotted line. (D) AFM topography and phase image of a DRG membrane (scalebar 1 µm). (E) AFM image of a hippocampal membrane showing three different levels: L0 glass, L1 single bilayer, L3 triple bilayer (folded). (F) the histogram reports the average membrane height and the standard error of the mean of n=10,000/3 z-positions/samples for each cell type (see Section II.G for the detailed procedure). }
	\label{fig:networks}
\end{figure*}

\subsection{Topographical characterization of native membranes at room temperature}

A natural instrument of investigation that operate in almost-physiological conditions is the AFM, but a non-trivial task is the localization of the region of interest when it is optically invisible. In the method presented here the localization is straightforward: in fact, the membrane is in the very same corner of the arrow (Figure 2A and 2B).
I performed AFM imaging of the intracellular side of the membrane of U87, U251, Hippocampal and DRG cells using cantilevers with spring constant of 0.08N/m in intermittent contact mode. The isolation method worked unexpectedly well with Hippocampal neurons, the AFM image (Figure 2C) revealed that the upper membrane is completely separated from the cell, and it covers the area where the cell got in contact with the arrow. The membrane surface is highly flat and it  displayed no cytoskeletal feature within the investigated dimensions (10 nm-10 $\mu$m range) unlike previously reported with other methods or cell types [10,14]. A higher resolution image indicate the presence of 30 nm-size complexes, most likely of proteic nature [16] (Figure S4). DRG’s membrane surfaces are qualitatively similar to Hippocampal ones, with the exception that the isolated membranes form island of \~2 $\mu$m2 (Figure 2D) instead of a continuous and complete layer of \~50 $\mu$m2. U87 and U251 membranes showed an intermediate behavior: continuous regions interspersed with more fragmented islands (Figure S4). The reasons of these different behaviors may be ascribed to a different membrane composition, or a different anchoring to the cytoskeleton that prevents a complete separation. Even with the same cell type there was some variability in the way the cell membrane is torn. For instance, the membrane near the border that did not adhere well to the polylysine, showed to favor a multi-bilayer conformation (Figure 2E) rather than floating freely in the solution. The dynamics of the membrane transfer is complicated, but it is clearly the behavior of a fluid layer that adapts to the shape of the support, differently  from a crystalline membrane of comparable thickness [32].
I perform also extensive characterization of the polylysine coated glass, which resulted to be affected by the violent separation of the membrane. In particular I tested the glass roughness (see Section II.G) in three conditions: i) before the isolation, ii) after the isolation far from the membrane, iii) after the isolation in proximity of the membrane. Condition (i) and (ii) didn’t show particular differences, while condition (iii) showed a 2-fold increase in the measured roughness. As suggested by Heuser [9], polylysine attracts the freely floating proteins in the solution. The hypothesis is that after the breakage of the cell, the cytosolic proteins diffuse around the site of the breakage, adhering to the coated glass and forming a layer on top of the polylysine. This hypothesis is supported by the Figure S4 which shows a portion of flat glass close to a region of more rough glass. The flat small region can be interpreted as an area that was shielded during the breakage. Therefore, the blob-like features on top of the membrane are presumably authentic and due to protein complexes, but they cannot be further resolved with the reported experimental conditions.  
I measured the membrane height of the four cell types according to the procedure explained in section II.G. The height distribution of the membrane in a single sample is typically Gaussian (Fig. S3). The height measures are then averaged over 3 samples: mean and standard error of the mean is reported in Figure 2F. The height of hippocampal and DRG membranes are of 8.2 nm and 8.3 nm respectively, they are \~2 nm thicker than the U87 (6.2 nm) and U251 (6.6 nm). The measured membrane heights are between 50\% and 100\% thicker than the well characterized model membranes dipalmitoylphosphatidylcholine (DPPC) or dioleoylphosphatidylcholine (DOPC) [24] that are 4-5 nm thick, and more similar to purple membrane [33]. 

\begin{figure*}
	
	\centering
	\includegraphics[width=0.6\linewidth]{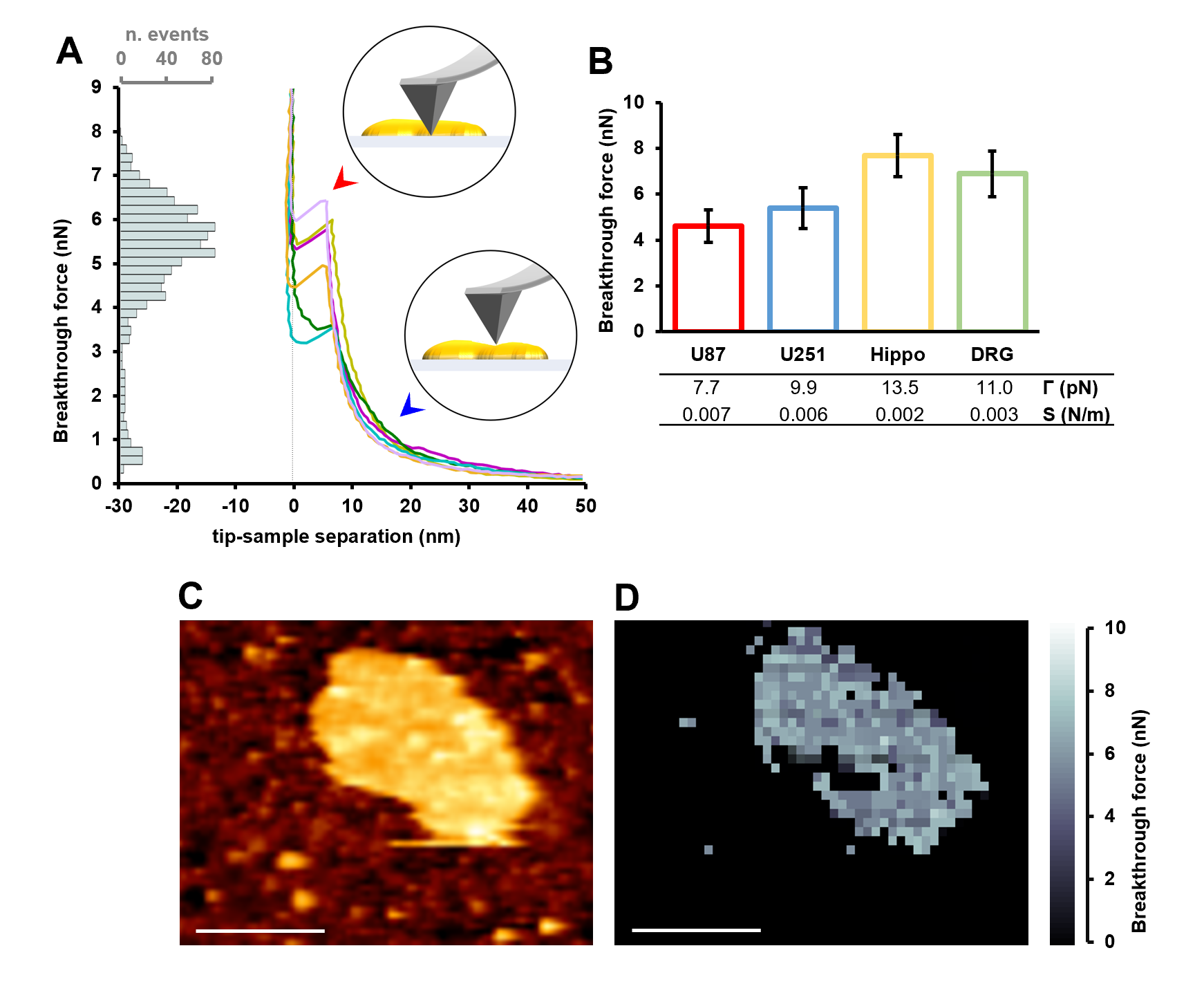}
	\captionsetup{justification=raggedright,singlelinecheck=false}
	\caption{(A) Force-distance indentation curves of U251 and the distribution of the rupture events. (B) average rupture forces and standard deviation of the distribution. The table presents the line tension $\Gamma$ and the spreading pressure S calculated with the nucleation model (see Section II.G). (C) AFM topographical image of DRG membrane before the (D) force map of the rupture events in the same region (scalebar 1 $\mu$m, 0 pN is intended for regions where no Breakthrough force was detected).}
	\label{fig:networks}
\end{figure*}

\subsection{Nanomechanical properties of the membranes }

After a precise determination of the position of the membrane, I could perform the puncturing of the bilayer, recording the resulting force-distance curves. The shape of the approach curves showed an initial elastic deformation (Figure 3A blue arrow) followed by an abrupt jump (5-8 nm; Figure 3A red arrow). The reversible compression of the bilayer only partially accounts for the reported elastic deformation, other contributions are probably due to some soft layer deposited on the tip and collected during scanning. The jump, on the other hand, clearly represents the membrane rupture that displays a step comparable with the height of the membrane.
I observed the presence of an additional rupture event in a non-negligible percentage of force curves (sometimes up to 20\%, Figure S5), which is due to an extra lipid bilayer formed on top of the cantilever tip [26]. The curves showing two rupture events were not considered in the quantitative analysis.
Figure 3B shows the average Breakthrough force of 800-1000 spectra for the four cell types and the standard deviation of the force distribution. To minimize the variability induced by the tip changes, new cantilevers were used in each experimental session. Hippocampal membranes revealed to pierce at force values distributed around 7.7 nN, that is almost twice the force necessary to pierce the U87 membranes which break at 4.6 nN. DRG and U251 pierce at values in between, 6.9 nN and 5.4 nN respectively. The full-width half maximum (FWHM) of the four force distributions is, on the other hand, more similar among cell types showing 20\% absolute variation at most. 
This isolation method in combination with a new module for the software Fodis [30] (see Section II.G), can generate also force-map images for a multiparametric characterization of the membranes (Figure 3C and 3D). The analysis that I performed at 24 Celsius revealed a single Gaussian distribution of the piercing forces for all the cell types, suggesting the presence of a unique liquid phase [27]. 
The Breakthrough force is just a proxy of the thermodynamic quantities of the lipid bilayer. For a more detailed analysis of the biophysical parameters of these native membranes, I applied the nucleation model [21] that was originally developed for synthetic preparations [27,31]. Thanks to the method presented here, the nucleation model can now be applied to native membranes, for quantifying the real line tension $\Gamma$ and spreading pressure S. The line tension $\Gamma$ is measured in pN and it indicates the energy per unit of length (i.e. the unsaturated bonds of the periphery of the membrane), while the spreading pressure S is the combination of the interfacial energies per unit of area of the membrane. In a configuration like the one presented here, where the pieces of membranes are exposed to the solution from one side and to the polylysine from the other, $\Gamma$ is the quantity that is minimized when the patches are circular (i.e. reducing the perimeter), while S is a measure of the tendency of spreading on the surface. The two neuron types have higher line tension compared to the two tumor cell lines but less than a half spreading pressure (values reported in Figure 3B), indicating that a higher line tension is related to a lower tendency to diffuse. These measurements provides values in the range of previous experiments [27,34], and particularly close to model membranes of mixtures of lipids (DOPC/sphingomyelin/cholesterol), but not to pure DOPC or DPPC which have 2-to-3 fold lower line tension and spreading pressure. Furthermore, this comparison shows that the experiments that involve model membranes should be performed preferentially with mixtures of lipids when the precondition is to mimic the nanomechanics of real cell membranes.

\section{Discussion}

I developed a new approach for imaging the cytoplasmic side of the cell membrane and for quantifying its nanomechanical properties. In this approach, the upper cell membrane is separated from the entire cell, therefore it is poised for various investigations. I used the AFM to characterize the topology and to probe the mechanics of these native preparations.
The approach was in this way applied to an endothelial cell type, two brain tumor cell lines and two neuron types, and it has proved to be successful for all the cells that attach well to the culture coverslip. Specific culturing protocols are required for those cells that only loosely bind to the substrate. 
Overall, these measurements are consistent with previous reports on synthetic membranes, but they provide additional information on the native state that are now easily accessible. 
The average membrane thickness reveals to be constant within the same cell type, but it differs by some nanometers over the tested cell types. The presence of extracellular matrix residues that may affect the height measure cannot be excluded a priori, but their presence in the upper part of the cultured cells should be modest if not negligible.  The height of the two tumor cell lines is similar to the well characterized purple membrane, and 25\% thinner than the neuron cell membrane.
The topological features are only partially corrupted by the substrate roughness, but to reach the goal of imaging diffusive single proteins in their native membrane it is necessary to use High speed-AFMs in serial dedicated investigations [35-37].
The Breakthrough force is higher in the neural cell types compared to the tumor cells, which suggests a higher rigidity of the membrane. This is in agreement with the measurements on synthetic lipid mixtures [28] and the higher cholesterol content of the neurons [38]. 
The single peak of the Breakthrough force distributions suggests the presence of a unique liquid phase in all the tested cells, as opposed to what has recently been shown in T-cells and HeLa cells where both the liquid ordered and liquid disordered phase was identified through fluorescence microscopy techniques [39]. It cannot be excluded that the AFM puncturing approach is not precise enough to resolve the coexistence of the two close phases (in terms of Breakthrough force) [28]. 
The nucleation model applied on native membranes allows to gather precise information on membrane packing, stability as well as intermolecular interactions among lipid molecules and proteins. The obtained values indicate that the neural membrane is more compact if compared to a more diffusive tumor membrane. In this regard, it is important to point out that the native membrane is deeply different from the model membranes that were tested in the past. Half of the weight of a native membrane is indeed due to membrane proteins [40]. The environment is highly diverse, therefore the biophysical parameters that was obtained here consider the overall contribution of lipids and proteins, leading to more robust values for computational lipidomics. To my knowledge, the sample preparation has always been problematic in previous unroofing techniques [13]. The single-cell method proposed here minimizes the sources of variability like large scale sonication or blotting, giving a success rate around 80\%.  The success rate becomes even more important when the unroofing needs to be operated ex/in vivo. To date, this method can directly be tuned for application on ex vivo thin tissues, but it is not immediately applicable in vivo. To overcome this problem, in the preparation process, the AFM should be substituted by a micromanipulator and the cells of interest should be accessible by the glass arrow. This kind of improvements could lead to an interesting tool for fast screening purposes of diseases that affect the cell membranes. 
There could be interesting applications also in complementary fields like single-molecule force spectroscopy (SMFS) or bioactive surface studies. Müller and collaborators has indeed recently found that we can expect a different behavior in the unfolding of proteins in a native environment [41], therefore this single-cell unroofing could be an ideal platform for such unfolding investigations. While having the possibility to deposit a real membrane on top of the desired artificial nanostructure could help to study some microscopic phenomena at the nanoscale, for instance, the antimicrobial activity of nanodots [42].
In conclusion, the method described here can be used to study membrane topology and mechanics at the nanoscale level, the scale at which important processes such as mechanosensing and membrane fusion take place, providing clues on their biophysical constraints.

\subsection*{Software Availability}
Fodis is available under the  Apache License, version 2.0. Source code, executables, datasets and full documentation are available for download in GitHub at \url{https://github.com/nicolagalvanetto/Fodis}. The latest release, version 1.2, is archived in Zenodo: DOI: 10.5281/zenodo.841277.

\subsection*{Acknowledgments}

The author tanks Prof Vincent Torre for the support and the suggestions. The author thanks Manuela Lough for checking the English. The author thanks Simone Mortal, Andrea Pedroni and Sourav Maity for the useful discussions.

This research did not receive any specific grant from funding agencies in the public, commercial, or not-for-profit sectors.

\bibliography{bib}

\subsection*{References}
[1]	J.D. Robertson, The ultrastructure of cell membranes and their derivatives, Biochem. Soc. Symp. 16 [1959] 3-43.\\

[2]	H. Fricke, The Electric Capacity of Suspensions with Special Reference to Blood, The Journal of General Physiology. 9 [1925] 137-152. doi:10.1085/jgp.9.2.137.\\

[3]	H. Davson, J.F. Danielli, The permeability of natural membranes., The Permeability of Natural Membranes. [1943]. https://www.cabdirect.org/cabdirect/abstract/19452200240 [accessed February 6, 2018].\\

[4]	M.V. Nermut, The ``cell monolayer technique`` in membrane research., Eur J Cell Biol. 28 [1982] 160-172.\\

[5]	M. Clarke, G. Schatten, D. Mazia, J.A. Spudich, Visualization of actin fibers associated with the cell membrane in amoebae of Dictyostelium discoideum, PNAS. 72 [1975] 1758-1762.\\

[6]	M.V. Nermut, L.D. Williams, Freeze-fracturing of monolayers [capillary layers] of cells, membranes and viruses: some technical considerations, Journal of Microscopy. 110 [1977] 121-132. doi:10.1111/j.1365-2818.1977.tb00023.x.\\

[7]	J.E. Heuser, T.S. Reese, M.J. Dennis, Y. Jan, L. Jan, L. Evans, Synaptic vesicle exocytosis captured by quick freezing and correlated with quantal transmitter release., The Journal of Cell Biology. 81 [1979] 275-300. doi:10.1083/jcb.81.2.275.\\

[8]	C.F. Peitsch, S. Beckmann, B. Zuber, iMEM: Isolation of Plasma Membrane for Cryoelectron Microscopy, Structure. 24 [2016] 2198-2206. doi:10.1016/j.str.2016.09.016.\\

[9]	J. Heuser, The Production of ``Cell Cortices`` for Light and Electron Microscopy, Traffic. 1 [2000] 545-552. doi:10.1034/j.1600-0854.2000.010704.x.\\

[10]	N. Morone, T. Fujiwara, K. Murase, R.S. Kasai, H. Ike, S. Yuasa, J. Usukura, A. Kusumi, Three-dimensional reconstruction of the membrane skeleton at the plasma membrane interface by electron tomography, J Cell Biol. 174 [2006] 851-862. doi:10.1083/jcb.200606007.\\

[11]	K.A. Sochacki, G. Shtengel, S.B. van Engelenburg, H.F. Hess, J.W. Taraska, Correlative super-resolution fluorescence and metal-replica transmission electron microscopy, Nature Methods. 11 [2014] nmeth.2816. doi:10.1038/nmeth.2816.\\

[12]	J. Usukura, A. Yoshimura, S. Minakata, D. Youn, J. Ahn, S.-J. Cho, Use of the unroofing technique for atomic force microscopic imaging of the intra-cellular cytoskeleton under aqueous conditions, J Electron Microsc [Tokyo]. 61 [2012] 321-326. doi:10.1093/jmicro/dfs055.\\

[13]	E. Usukura, A. Narita, A. Yagi, S. Ito, J. Usukura, An Unroofing Method to Observe the Cytoskeleton Directly at Molecular Resolution Using Atomic Force Microscopy, Scientific Reports. 6 [2016] srep27472. doi:10.1038/srep27472.\\

[14]	F. Sato, H. Asakawa, T. Fukuma, S. Terada, Semi- in situ atomic force microscopy imaging of intracellular neurofilaments under physiological conditions through the ``sandwich`` method, Microscopy [Oxf]. 65 [2016] 316-324. doi:10.1093/jmicro/dfw006.\\

[15]	S. Janel, E. Werkmeister, A. Bongiovanni, F. Lafont, N. Barois, Chapter 9 - CLAFEM: Correlative light atomic force electron microscopy, in: T. Müller-Reichert, P. Verkade [Eds.], Methods in Cell Biology, Academic Press, 2017: pp. 165-185. doi:10.1016/bs.mcb.2017.03.010.\\

[16]	L. Picas, F. Rico, M. Deforet, S. Scheuring, Structural and Mechanical Heterogeneity of the Erythrocyte Membrane Reveals Hallmarks of Membrane Stability, ACS Nano. 7 [2013] 1054-1063. doi:10.1021/nn303824j.\\

[17]	E.M. Schmid, M.H. Bakalar, K. Choudhuri, J. Weichsel, H.S. Ann, P.L. Geissler, M.L. Dustin, D.A. Fletcher, Size-dependent protein segregation at membrane interfaces, Nature Physics. 12 [2016] 704-711. doi:10.1038/nphys3678.\\

[18]	T. Bhatia, J. Agudo-Canalejo, R. Dimova, R. Lipowsky, Membrane Nanotubes Increase the Robustness of Giant Vesicles, ACS Nano. [2018]. doi:10.1021/acsnano.8b00640.\\

[19]	Z. Al-Rekabi, S. Contera, Multifrequency AFM reveals lipid membrane mechanical properties and the effect of cholesterol in modulating viscoelasticity, PNAS. 115 [2018] 2658-2663. doi:10.1073/pnas.1719065115.\\

[20]	Y.F. Dufrêne, T. Boland, J.W. Schneider, W.R. Barger, G.U. Lee, Characterization of the physical properties of model biomembranes at the nanometer scale with the atomic force microscope, Faraday Discussions. 111 [1999] 79-94. doi:10.1039/A807637E.\\

[21]	S. Loi, G. Sun, V. Franz, H.-J. Butt, Rupture of molecular thin films observed in atomic force microscopy. II. Experiment, Phys. Rev. E. 66 [2002] 031602. doi:10.1103/PhysRevE.66.031602.\\

[22]	R.P. Gonçalves, G. Agnus, P. Sens, C. Houssin, B. Bartenlian, S. Scheuring, Two-chamber AFM: probing membrane proteins separating two aqueous compartments, Nature Methods. 3 [2006] nmeth965. doi:10.1038/nmeth965.\\

[23]	S. Garcia-Manyes, F. Sanz, Nanomechanics of lipid bilayers by force spectroscopy with AFM: A perspective, Biochimica et Biophysica Acta [BBA] - Biomembranes. 1798 [2010] 741-749. doi:10.1016/j.bbamem.2009.12.019.\\

[24]	Z.V. Leonenko, E. Finot, H. Ma, T.E.S. Dahms, D.T. Cramb, Investigation of Temperature-Induced Phase Transitions in DOPC and DPPC Phospholipid Bilayers Using Temperature-Controlled Scanning Force Microscopy, Biophysical Journal. 86 [2004] 3783-3793. doi:10.1529/biophysj.103.036681.\\

[25]	U. Bhojoo, M. Chen, S. Zou, Temperature induced lipid membrane restructuring and changes in nanomechanics, Biochimica et Biophysica Acta [BBA] - Biomembranes. 1860 [2018] 700-709. doi:10.1016/j.bbamem.2017.12.008.\\

[26]	S. Garcia-Manyes, G. Oncins, F. Sanz, Effect of Ion-Binding and Chemical Phospholipid Structure on the Nanomechanics of Lipid Bilayers Studied by Force Spectroscopy, Biophysical Journal. 89 [2005] 1812-1826. doi:10.1529/biophysj.105.064030.\\

[27]	S. Chiantia, J. Ries, N. Kahya, P. Schwille, Combined AFM and Two-Focus SFCS Study of Raft-Exhibiting Model Membranes, ChemPhysChem. 7 [2006] 2409-2418. doi:10.1002/cphc.200600464.\\

[28]	R.M.A. Sullan, J.K. Li, C. Hao, G.C. Walker, S. Zou, Cholesterol-Dependent Nanomechanical Stability of Phase-Segregated Multicomponent Lipid Bilayers, Biophysical Journal. 99 [2010] 507-516. doi:10.1016/j.bpj.2010.04.044.\\

[29]	D. Nečas, P. Klapetek, Gwyddion: an open-source software for SPM data analysis, Open Physics. 10 [2012] 181-188. doi:10.2478/s11534-011-0096-2.\\

[30]	N. Galvanetto, A. Perissinotto, A. Pedroni, V. Torre, Fodis: Software for Protein Unfolding Analysis, Biophysical Journal. 114 [2018] 1264-1266. doi:10.1016/j.bpj.2018.02.004.\\

[31]	A.J. García-Sáez, S. Chiantia, P. Schwille, Effect of Line Tension on the Lateral Organization of Lipid Membranes, J. Biol. Chem. 282 [2007] 33537-33544. doi:10.1074/jbc.M706162200.\\

[32]	K. Schouteden, N. Galvanetto, C.D. Wang, Z. Li, C. Van Haesendonck, Scanning probe microscopy study of chemical vapor deposition grown graphene transferred to Au[111], Carbon. 95 [2015] 318-322. doi:10.1016/j.carbon.2015.08.033.\\

[33]	M. Stark, C. Möller, D.J. Müller, R. Guckenberger, From Images to Interactions: High-Resolution Phase Imaging in Tapping-Mode Atomic Force Microscopy, Biophysical Journal. 80 [2001] 3009-3018. doi:10.1016/S0006-3495[01]76266-2.\\

[34]	J.D. Moroz, P. Nelson, Dynamically stabilized pores in bilayer membranes, Biophysical Journal. 72 [1997] 2211-2216. doi:10.1016/S0006-3495[97]78864-7.\\

[35]	I. Casuso, J. Khao, M. Chami, P. Paul-Gilloteaux, M. Husain, J.-P. Duneau, H. Stahlberg, J.N. Sturgis, S. Scheuring, Characterization of the motion of membrane proteins using high-speed atomic force microscopy, Nature Nanotechnology. 7 [2012] 525-529. doi:10.1038/nnano.2012.109.\\

[36]	J. Preiner, A. Horner, A. Karner, N. Ollinger, C. Siligan, P. Pohl, P. Hinterdorfer, High-Speed AFM Images of Thermal Motion Provide Stiffness Map of Interfacial Membrane Protein Moieties, [2014]. doi:10.1021/nl504478f.\\

[37]	Y. Ruan, K. Kao, S. Lefebvre, A. Marchesi, P.-J. Corringer, R.K. Hite, S. Scheuring, Structural titration of receptor ion channel GLIC gating by HS-AFM, Proceedings of the National Academy of Sciences. [2018] 201805621. doi:10.1073/pnas.1805621115.\\

[38]	H.I. Ingólfsson, T.S. Carpenter, H. Bhatia, P.-T. Bremer, S.J. Marrink, F.C. Lightstone, Computational Lipidomics of the Neuronal Plasma Membrane, Biophysical Journal. 113 [2017] 2271-2280. doi:10.1016/j.bpj.2017.10.017.\\

[39]	D.M. Owen, D.J. Williamson, A. Magenau, K. Gaus, Sub-resolution lipid domains exist in the plasma membrane and regulate protein diffusion and distribution, Nature Communications. 3 [2012]. doi:10.1038/ncomms2273.\\

[40]	G.M. Cooper, The Cell: A Molecular Approach, 2nd ed., 2000.\\

[41]	J. Thoma, S. Manioglu, D. Kalbermatter, P.D. Bosshart, D. Fotiadis, D.J. Müller, Protein-enriched outer membrane vesicles as a native platform for outer membrane protein studies, Communications Biology. 1 [2018] 23. doi:10.1038/s42003-018-0027-5.\\

[42]	G. Benetti, E. Cavaliere, A. Canteri, G. Landini, G.M. Rossolini, L. Pallecchi, M. Chiodi, M.J. Van Bael, N. Winckelmans, S. Bals, L. Gavioli, Direct synthesis of antimicrobial coatings based on tailored bi-elemental nanoparticles, APL Materials. 5 [2017] 036105. doi:10.1063/1.4978772.\\

\balancecolsandclearpage

\begin{figure*}
	
	\centering
	\includegraphics[width=0.8\linewidth]{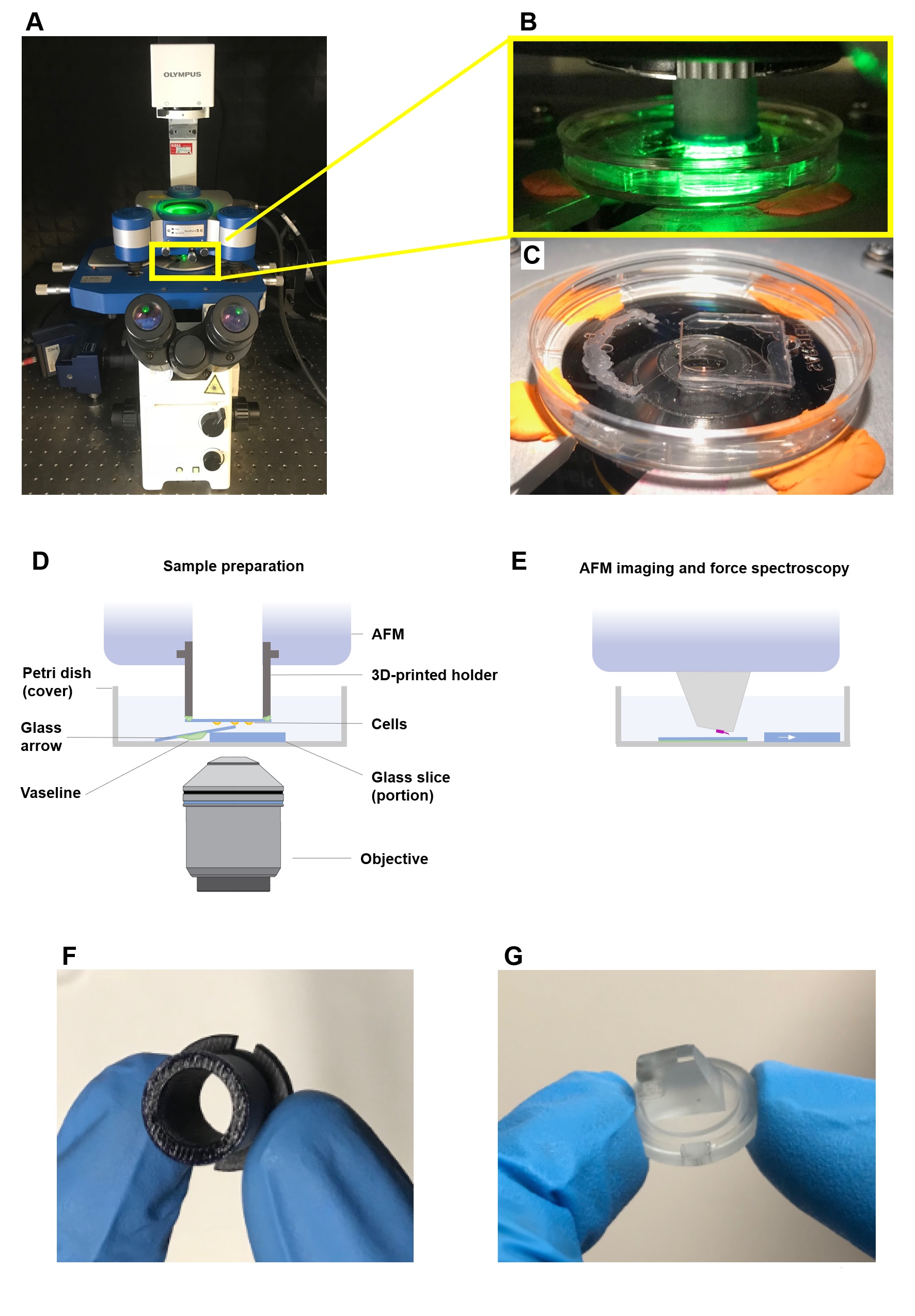}
	\captionsetup{justification=raggedright,singlelinecheck=false}.
	\makeatletter
	\renewcommand{\fnum@figure}{\figurename~S1}
	\makeatother
	\caption{\textbf{Details of the Setup}:  (A) photo of the setup used: the Olympus microscope and the JPK AFM. (B) magnification of the components drawn in (D). (C)  configuration of the tilted glass arrow and glass slice in (D). (D) scheme of the experiment during compression of the cells. (E) scheme of the configuration during AFM imaging or Force Spectroscopy. (F) 3D printed cell culture holder in comparison with (G) the cantilever holder.  }
	
\end{figure*}

\begin{figure*}
	
	\centering
	\includegraphics[width=0.7\linewidth]{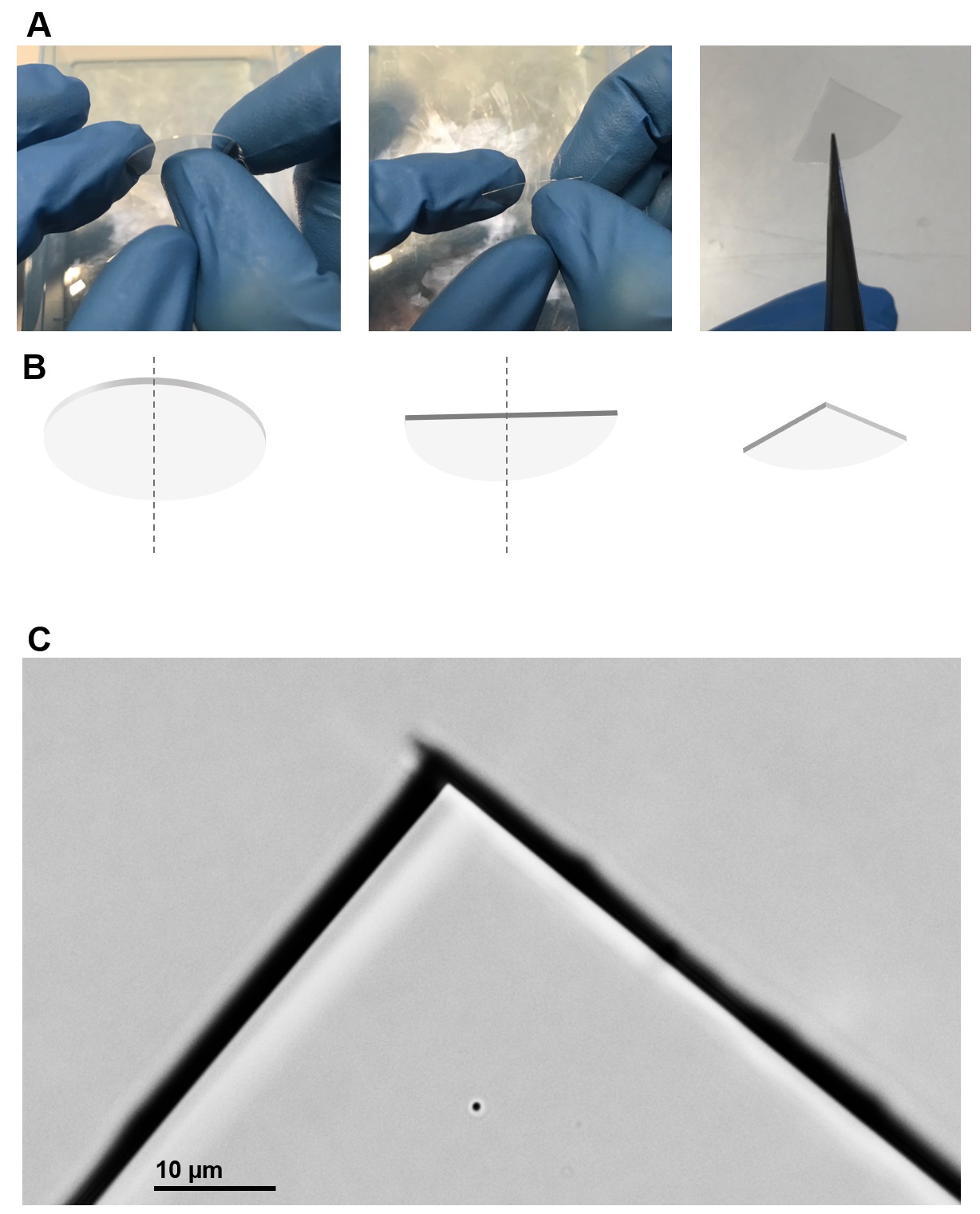}
	\captionsetup{justification=raggedright,singlelinecheck=false}.
	\makeatletter
	\renewcommand{\fnum@figure}{\figurename~S2}
	\makeatother
	\caption{\textbf{Glass Arrows Preparation}:  (A) the 3 steps for the preparation of the glass arrow from a 24 mm coverslip (see Materials and Methods). Step 1 ((A) left): the round coverslip is broken in 2 pieces pushing with the thumb in the middle of the coverslip. Step 2 ((A) center): the half-coverslip is broken in other 2 pieces pushing with the thumb in the middle of the half-coverslip. Step 3 ((A) right): the sharpness of the apex is checked with the microscope (C). (B) scheme of the coverslip stages during the preparation in (A). (C) the apex of the glass arrow results to be optically sharp, as opposed to a cut with the typical diamond pen that shows 5 $\mu$m-size scratches along the edge.
		}
	
\end{figure*}

\begin{figure*}
	
	\centering
	\includegraphics[width=0.8\linewidth]{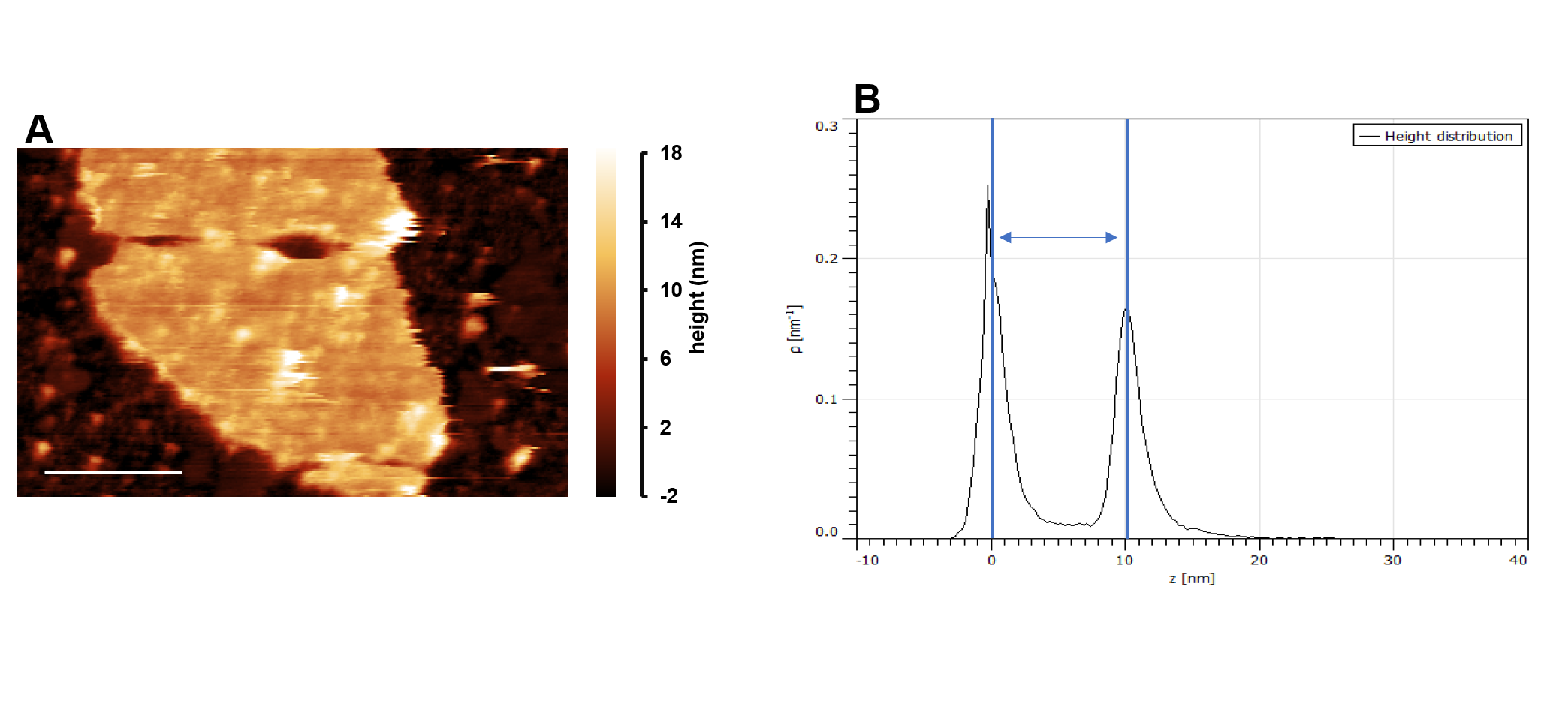}
	\captionsetup{justification=raggedright,singlelinecheck=false}.
	\makeatletter
	\renewcommand{\fnum@figure}{\figurename~S3}
	\makeatother
	\caption{\textbf{Height Measurements}:  (A) AFM image (scalebar 1 $\mu$m) and corresponding (B) height distribution in Gwyddion. The reported height in Figure 2F is the difference between the means of the two bells.}
	
\end{figure*}

\begin{figure*}
	
	\centering
	\includegraphics[width=0.6\linewidth]{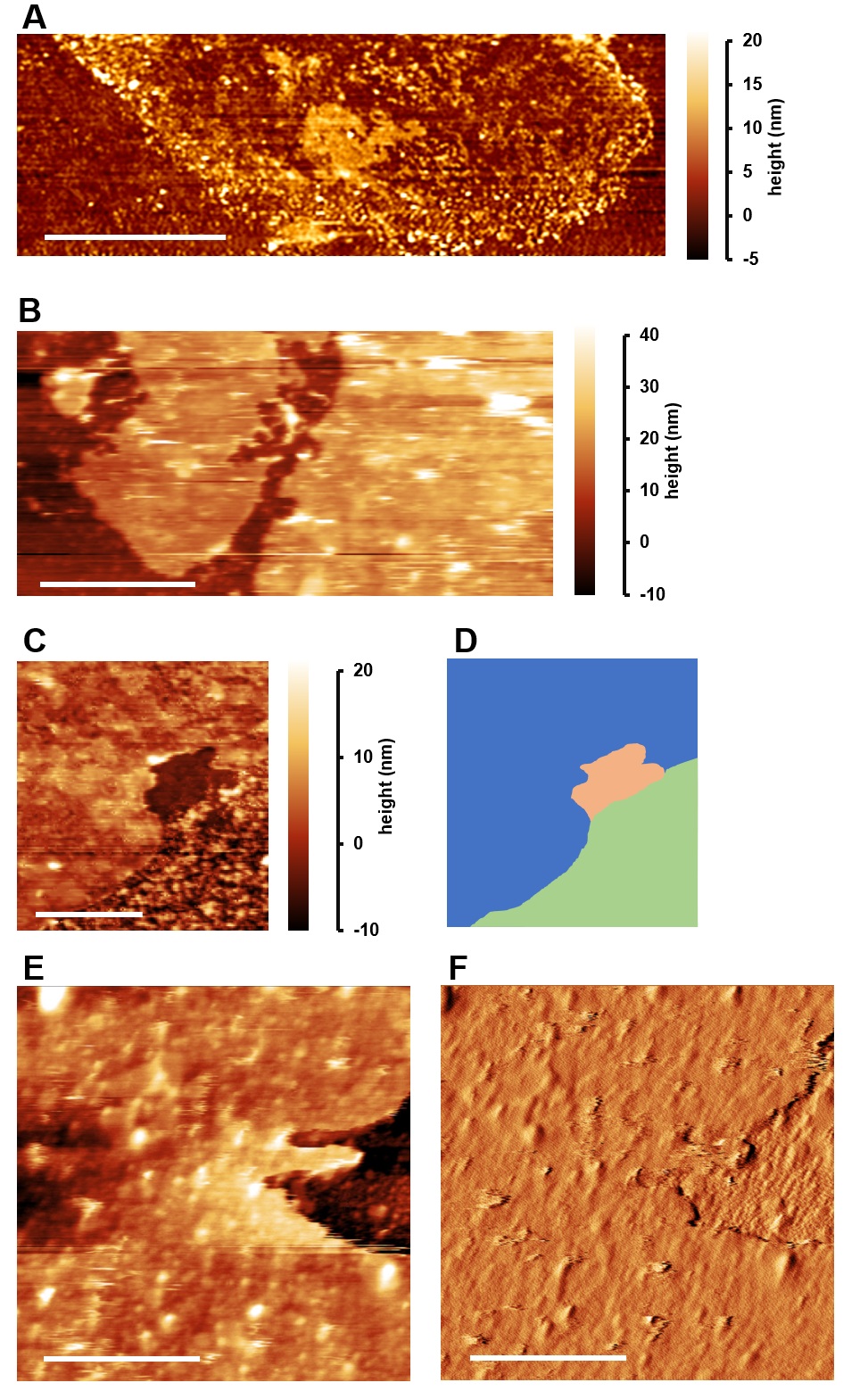}
	\captionsetup{justification=raggedright,singlelinecheck=false}.
	\makeatletter
	\renewcommand{\fnum@figure}{\figurename~S4}
	\makeatother
	\caption{\textbf{AFM Imaging}: (A) AFM image of a U251 cell membrane (scale bar 4 $\mu$m). (B) AFM image of a U87 cell membrane (scale bar 4.5 $\mu$m). (C) AFM image of a Hippocampal cell membrane (scale bar 2 $\mu$m) displaying three different surface features (D): (blue) surface of the membrane, (green) rough glass and (orange) smooth glass. I interpret the roughness in the green portion as the deposition of the cytosolic proteins of the broken cell, the orange portion was instead shielded during the unroofing process. (E) High resolution AFM image and (F) error signal of a Hippocampal membrane border (scale bar 500 nm). }
	
\end{figure*}

\begin{figure*}
	
	\centering
	\includegraphics[width=0.4\linewidth]{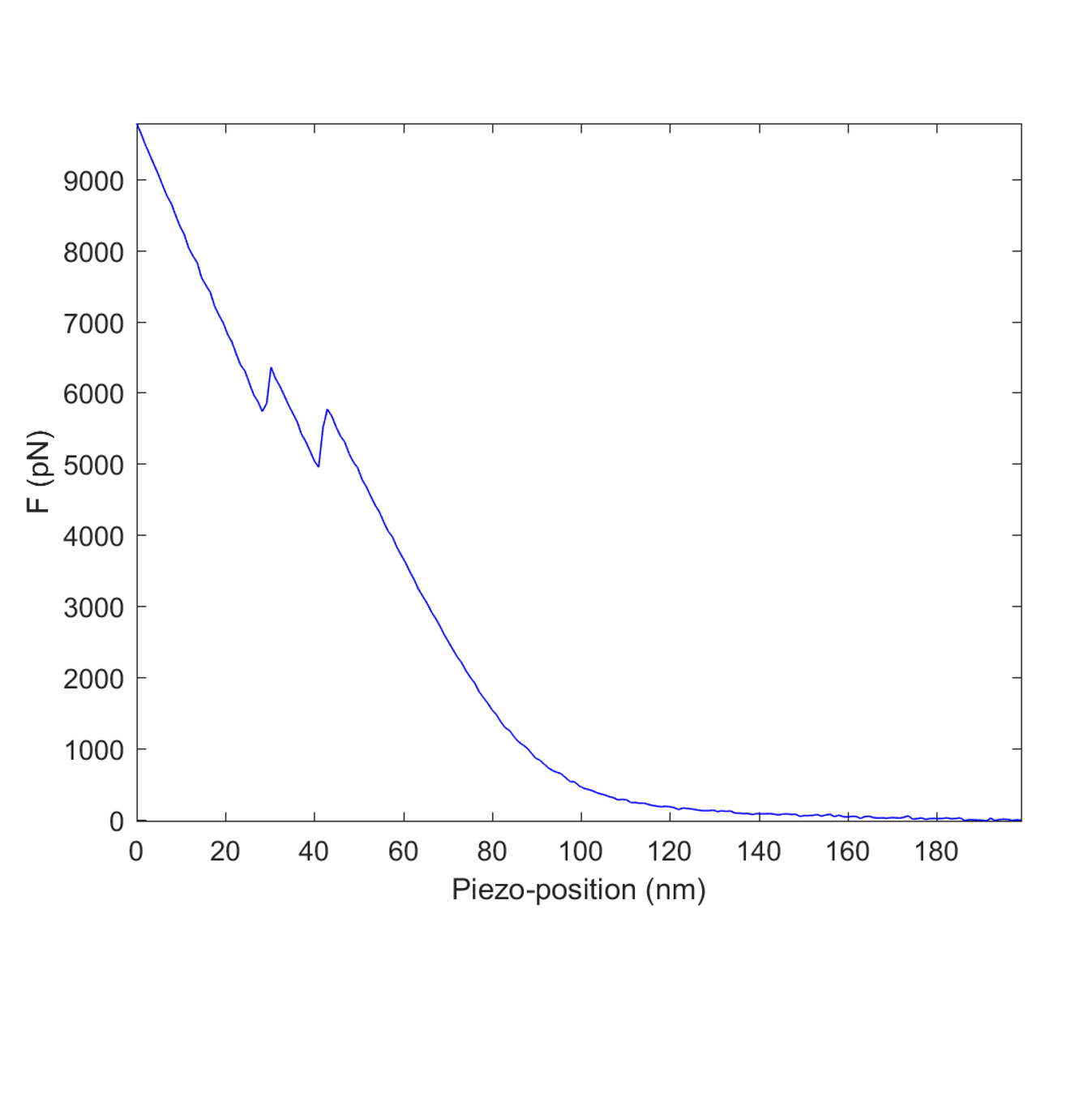}
	\captionsetup{justification=raggedright,singlelinecheck=false}.
	\makeatletter
	\renewcommand{\fnum@figure}{\figurename~S5}
	\makeatother
	\caption{\textbf{Two-step Spectrum}:  Force-distance curve showing a two-step curve that indicate the presence of an extra bilayer of lipids. }
	
\end{figure*}

\balancecolsandclearpage

\end{document}